\documentclass[10pt,leqno]{amsart}
\usepackage{graphicx}
\usepackage{indentfirst,csquotes}
\usepackage{amssymb,amsthm,amsmath}
\usepackage{xcolor,paralist,hyperref,titlesec,fancyhdr,etoolbox}
\usepackage{multirow}

\titleformat{\section}[display]{\normalfont\huge\bfseries\centering}{\centering\chaptertitlename\thechapter}{10pt}{\Large}
\titlespacing*{\section}{0pt}{0ex}{0ex}

\hypersetup{ colorlinks=true, linkcolor=black, filecolor=black, urlcolor=black }

\usepackage{lipsum}

\begin{document}

\title{Design Analysis and Experimental Validation of Relaxation Oscillator-Based Circuit for R-C Sensors} 

\author{Mohamad Idris Wani$^1$, Sadan Saquib Khan$^1$, Benish Jan$^1$, Meraj Ahmad$^2$, Maryam Shojaei Baghini$^3$, Laxmeesha Somappa$^3$, and Shahid Malik$^1$}
\address{$^1$Centre for Sensors, Instrumentation, and Cyber Physical System Engineering, Indian Institute of Technology, New Delhi, India}
\address{$^2$James Watt School of Engineering, University of Glasgow, United Kingdom}
\address{$^3$Department of Electrical Engineering, Indian Institute of Technology Bombay, Mumbai, India}
\email{smalik@iitd.ac.in}

\maketitle


\begin{abstract}
 Relaxation oscillator-based circuits are widely used for interfacing various resistive and capacitive sensors. The electrical equivalent of most resistive and capacitive sensors is represented using a parallel combination of resistor and capacitor. The relaxation oscillator-based circuits are not suitable for parallel R-C sensors. In this paper, we propose a modified circuit for parallel R-C sensors. The proposed relaxation oscillator-based circuit is based on a dual-slope and charge transfer technique to measure the resistance and capacitance of parallel R-C sensors separately. In addition, the paper provides a detailed analysis and design considerations for the oscillator design by taking into account the various sources of non-idealities. A method to reduce the error by using single-cycle averaging is also introduced. To verify the analyzed design criteria, the circuit is tested with multiple operational amplifiers with different non-idealities. Experimental results verify the performance of the proposed circuit. The circuit is tested for a range from 10 pF to 42 pF and 100 k$\Omega$ to 1 M$\Omega$ for parallel R-C sensors with an error of less than 1.5\%. The circuit is tested with a fabricated water-level sensor. The result confirms the efficacy of the proposed circuit.
\end{abstract} 

\bigskip

\textbf{Introduction}\\
{R}{esistive} and capacitive sensors are very popular and widely used in the scientific, biomedical, and industrial applications for measurement of force, displacement, pressure, humidity, temperature, liquid level, and bio-medical implants \cite{1,2,3,4,5}. The signal conditioning circuits for these sensors are generally based on bridge-based circuits, relaxation oscillator-based resistance/capacitance to frequency converters, and direct digital converters \cite{6,7,8,9}.\\
Bridge circuits are the fundamental building block for resistance measurement. Wheatstone bridge-based circuits are preferred for measuring a small variation in sensor resistance. However, the single-element resistive sensors' output voltage is non-linear with respect to the change in the sensor resistance \cite{10}. This is especially critical when the sensor resistance change is large compared to the baseline resistance. This affects the sensitivity for wide dynamic range measurement. On the other hand, the relaxation oscillator-based circuits provide a wide-dynamic range measurement of sensor resistance by linearly converting the resistance into frequency. \\ 

Most of the signal conditioning circuits for resistive and capacitive sensors are based on the assumption that the sensor is ideal and can be represented by either a resistor in the case of the resistive sensors or a capacitor in the case of the capacitive sensors \cite{6,9,11,12,13}. However, the assumption is not valid for many sensing applications. For instance, the resistive sensors are affected by a parasitic capacitance \cite{2}. Similarly, most of the capacitive sensors are affected by a leakage resistance \cite{14,15}. Therefore, the electrical equivalent circuit for most of the resistive and capacitive sensors consists of a resistor and capacitor in parallel \cite{2}. In such sensors, one component carries information about the measurand, while the other component is present because of the non-ideal nature of the sensor which is unpredictable. Therefore, its effect should be compensated at the circuit level, or it degrades the performance of the sensor system. Moreover, in some sensors, both sensing components are a function of the measurand and need simultaneous measurement \cite{16,17}.

The signal conditioning circuits for the parallel R-C sensors are mostly based on the phase-sensitive detection (PSD) technique. The PSD-based techniques require an in-phase and a quadrature reference signal to separately measure the resistance and capacitance of the parallel R-C sensors. Auto-nulling and PSD-based signal conditioning circuits for the parallel R-C sensors are reported in \cite{18,19,20}. These circuits periodically provide the measurement of the R-C parameters of the sensor. However, it requires analog multipliers and an auto-nulling loop, which increases the complexity and power consumption. Another PSD-based signal conditioning circuit for parallel R-C sensors is reported in \cite{21}. The circuit provides simultaneous measurement of sensor capacitance and resistance. However, the output of the signal conditioning circuit is sensitive to many circuit parameters and component non-idealities, which affects the robustness of the sensor system. Moreover, the output voltage is sensitive to noise, affecting the measurement's resolution.

\begin{figure}[h]
    \centering
    \includegraphics[trim=5 15 5 15, clip, scale = 1.2]{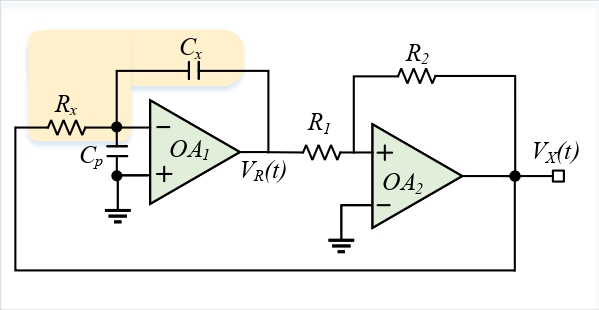}
     \caption{Schematic diagram of the  relaxation oscillator circuit for resistive/capacitive sensor. In the case of an ideal capacitive sensor, $C_x$ is the sensing variable, and $R_x$ is the integrator resistance. In the case of an ideal resistive sensor, $R_x$ is the sensing element, and $C_x$ is the integrator capacitance as a reference element.}   
     \label{fig:circuit_1}
\end{figure}

Thanks to the inherent properties such as the quasi-digital output and high noise immunity, the relaxation oscillator-based circuits are preferred for lossless resistive, and capacitive sensors \cite{7,8,22,23} as shown in Fig. \ref{fig:circuit_1}. The oscillator-based circuit reported in \cite{23,24,25} are designed for the measurement of larger variation in sensor resistance with parasitic capacitance. However, the design considerations for the desired accuracy are not presented in \cite{23,24}. Another oscillator-based signal conditioning circuit for parallel R-C sensors is reported in \cite{14}; however, the measurement range for the capacitance measurement is very small ($<$1 pF). This limits the range of applications of the circuit reported in \cite{14}.

The initial results of the relaxation-oscillator circuit for capacitance measurement are reported in \cite{25}. In this paper, we present the extended analysis, design considerations, experimental validation of the performance parameters, and sensor testing with the relaxation oscillator-based signal conditioning circuits. The features of the paper are highlighted as follows. 

\begin{itemize}
\item  The proposed oscillator combines dual-slope and charge-transfer techniques to separately measure the resistance and capacitance of the parallel R-C sensors. 
\item Detailed analysis, including various sources of non-idealities, is included in the paper. The effect of various component non-idealities on the measurement of sensor parameters is analyzed. 
\item A single-cycle averaging method is utilized, which significantly reduces the effects of component non-idealities on the measurement of sensor parameters. 
\item The design criteria for the selection of components for designing the proposed signal conditioning circuit are also included in the paper. 

\item The proposed circuit is tested for different sensing conditions and the measurement is compared with the derived analytical solution.
\end{itemize}

\textbf{Conventional Relaxation Oscillator for Resistive--Capacitive Sensors}\\
The schematic diagram of the relaxation oscillator-based signal conditioning circuit (with triangular and square wave output) for resistive and capacitive sensors is shown in Fig. \ref{fig:circuit_1}. The circuit consists of an integrator and a Schmitt trigger. The operational amplifier $OA_1$ along with the resistance $R_x$ and capacitor $C_x$ form the integrator of the oscillator. In the case of a capacitive sensor, $C_x$ is replaced by the sensor element, and $R_x$ is implemented using a known resistor. Similarly, in the case of resistive sensors, $R_x$ is the sensor element (considered to be lossless in Fig.\ref{fig:circuit_1}), and $C_x$ is the reference element. The signal conditioning circuits based on the conventional relaxation oscillator for resistive and capacitive sensors are reported in \cite{7,8,22}.

The integrator output in Fig. \ref{fig:circuit_1} is compared with the threshold voltage of the Schmitt trigger (implemented using $R_1$, $R_2$, and Op-Amp $OA_2$). The expression for the voltage $V_R(t)$ can then be derived as follows. 

\begin{figure}[t]
    \centering
    \includegraphics[trim=15 18 15 18, clip, scale = 1.2]{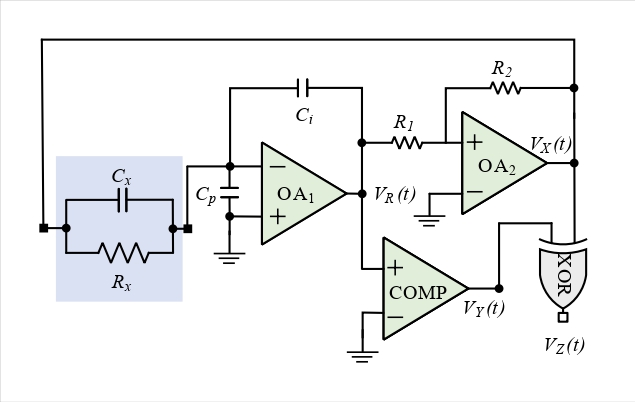}
     \caption{ Schematic diagram of the proposed modified relaxation oscillator for parallel R-C sensors}   
     \label{fig:circuit_4}
\end{figure}
\begin{equation}
   V_R(t) = - \frac{1}{R_x C_x} \; \int_{0}^t V_p \; dt
    \label{eq:V_R(t)_conventional}
\end{equation}
where $V_p$ is the peak voltage at the output of the Schmitt trigger. The threshold voltage of the Schmitt trigger is given as $\pm V_p \times (R_1/R_2)$. The capacitor $C_x$ is initially charged with voltage ${|V_p|}$ at $t=0$. At $t=T/2$, the voltage $V_R(t)$ can be written as follows. 
\begin{equation}
    V_R\left(t=\frac{T}{2}\right) =- \frac{T \; V_p }{2\; R_x \; C_x} =   - 2\; V_p \; \frac{R_1}{R_2}
\end{equation}
The expression of the oscillation time-period $T$ of the square-wave output signal $V_X(t)$ can be derived as follows. 
\begin{equation}
  T= 4\;R_x \; C_x \; \frac{R_1}{R_2}
  \label{T_conventional}
\end{equation}

The expression (\ref{T_conventional}) indicates that the period $T$ of the square-wave output of the relaxation oscillator circuit is proportional to the sensor variables (either $C_x$ or $R_x$ at once). However, the conventional relaxation oscillator circuit in Fig. \ref{fig:circuit_1} is not suitable for the leaky capacitive and parallel R-C sensors.\\

The accuracy of the relaxation oscillator based circuits for sensing applications are highly dependent on the symmetrical output in both positive and negative polarity. This can either be achieved using a network of Zener diodes or by using a rail-to-rail operational amplifier for the Schmitt trigger. Further, a current limiting resistor can also be used to ensure the safety \cite{26}. In this paper, we are assuming the output voltage of the Schmitt trigger is symmetrical.

\textbf{Modified Relaxation Oscillator for Impedance R-C Sensors}\\
The modified relaxation oscillator circuit with the parallel $R_x$ and $C_x$ (electrical equivalent of a resistive sensor) at the inverting terminal of the integrator is shown in Fig. \ref{fig:circuit_4}. The capacitor $C_x$ is active only during the switching of square-wave voltage $V_X(t)$ from low to high and vice-versa. {Charge transfer occurs between the capacitors $C_x$ and $C_i$, during each of these transitions}. For the rest of the cycle, the capacitor $C_i$ {charges/discharges} through $R_x$ with a linear slope. {$C_p$ is the parasitic capacitance of op-amp}. 
\par

\textbf{Oscillator Operating Principle}
The expression for the voltage $V_R(t)$ {in Fig.~\ref{fig:circuit_4}} can be derived as follows.  
\begin{equation}
    V_R(t) = \left(\frac{R_1}{R_2} - \frac{2\;C_x}{C_i} \right)V_p - \int_{0}^{t} \frac{V_p}{R_{x} \; C_i} dt
\end{equation}
The factor $2C_x/C_i$ is due to the charge transfer between $C_x$ and $C_i$. The sudden charge transfer during the transition affects the initial voltage at the output $V_R(t)$. At $t= T/2$, $V_R(t)$ can be simplified as follows. 
\begin{equation}
    V_R(t) = \left(\frac{R_1}{R_2} - \frac{2\;C_x}{C_i} \right)V_p - \frac{T\;V_p}{2\;R_{x} \; C_i} = - V_p \frac{R_1}{R_2}
\end{equation}
Therefore, {the oscillation period, $T$ at the output node $V_x$ can be expressed as follows.}
\begin{equation}
   T = 4 \; R_x \; C_i \; \left( \frac{R_1}{R_2} - \frac{C_x}{C_i} \right)
\end{equation}
The oscillation period $T$ at the output node $V_x$ of the circuit in Fig. \ref{fig:circuit_4} is proportional to $R_x$ and $C_x$. However, the expression of time-period $T$ is not enough to separately estimate both the capacitance $C_x$ and resistance $R_x$ of the sensor. {Hence, additional information is necessary to measure both parameters separately.} 
\begin{figure}
    \centering
    \includegraphics[trim=22 15 15 15, clip, scale = 1.4]{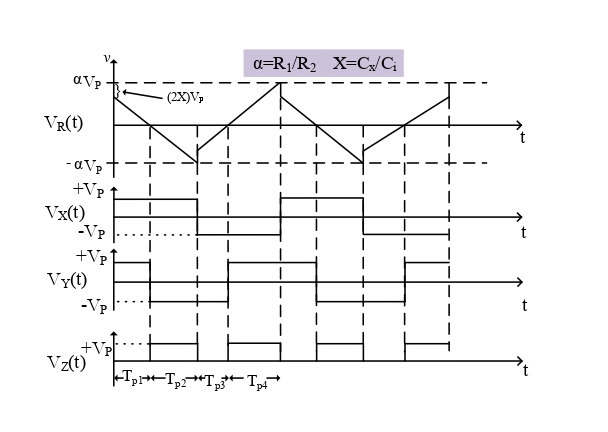}
     \caption{ Timing diagram of the proposed oscillator shown in Fig. \ref{fig:circuit_4}}   
     \label{fig:circuit_5}
\end{figure}

To measure both the parameters $R_x$ and $C_x$, a zero-crossing-detector ({ZCD}) (implemented using $OA_3$) and an XOR gate is added to separate the sensor components ($R_x$ \& $C_x$) into an equivalent pulse width. The ZCD separates the ramp signal into two parts: The first part is proportional to the charge transfer, and the second part is proportional to sensor resistance as represented by $T_{p1}$ and $T_{p2}$ in Fig. \ref{fig:circuit_5}. The expression for $T_{p1}$ and $T_{p2}$ can be derived as follows.
\begin{equation}
    T_{p1} = R_x \; C_i\left( \frac{R_1}{R_2} - \frac{2 \; C_x}{C_i} \right)
\end{equation}
\begin{equation}
   T_{p2} = R_x\;  C_i \; \frac{R_1}{R_2}
\end{equation}

\noindent The expression of unknown impedance $(R_x, C_x)$ can be estimated as from $T_{p1}$ and $T_{p2}$ as follows. 

\begin{equation}
   C_x =  \frac{R_1}{R_2} \frac{C_i}{2} (1-\frac{T_{p1}}{T_{p2}})
   \label{eq:capacitance_eq}
\end{equation}

\begin{equation}
   R_x =  \frac{R_1}{C_i R_2} T_{p2}
\end{equation}

\textbf{Oscillator Performance Analysis with Component Non-idealities}\\
The developed relaxation oscillator circuit's output periods are affected by many circuit parameters and must be analyzed to design the circuit for a desired target resolution. The following non-idealities are considered.
\begin{itemize}
    \item Op-Amp finite GBW, $\mathrm{A_0\omega_0}$
    \item Op-Amp input bias current, $\mathrm{i_b}$
    \item Op-Amp slew rate, SR
    \item Schmitt trigger and Op-Amp offset voltage, $\mathrm{V_{os}}$
    \item ZCD comparator offset, $\mathrm{V_{oz}}$
    \item ZCD comparator response delay, $\mathrm{\tau_{Z,LH},\tau_{Z,HL}}$
    \item Schmitt trigger response delay, $\mathrm{\tau_{S,LH},\tau_{S,HL}}$
\end{itemize}

The propagation time of the XOR gate will also affect the periods. The parasitic capacitance of the XOR gate also contributes to the propagation time. Usually, the propagation time of the XOR gate is in ns, which is smaller than the zero crossing detector and Schmitt trigger used in the circuit. Therefore, the effect of the XOR gate can be considered negligible for the circuit. Further, the effect of the propagation time of the XOR gate can be considered in the equation by adding the rise and fall time with the comparator output.

The steps for analyzing the effect of component nonidealities on the measurement of sensor parameters are as
follows.
\begin{enumerate}
    \item First, we will analyze the effect of the op-amp nonidealities on the slope and the offset of the integrator. The change in the integration slope and the effective integrator output offset is the primary concern for the oscillators' performance.
    \item Next, we consider the effect of the oscillation period due to the component non-idealities in the ZCD and the Schmitt trigger with the non-ideal integrator output applied.
    \item Finally, we will combine all the non-idealities to derive
the oscillation period at the output of the XOR gate.
\end{enumerate}

The detailed analysis of the component non-idealities is as
follows.\\
\textbf{ Effect of the op-amp non-idealities on the slope and the
offset of the integrator}

First, considering finite GBW product, the integrator output voltage when driven by a unit step voltage $\mathrm{V_p}$ can be derived as follows.
\begin{align}
    V_R(s) &= \frac{-V_p A_0 \omega_0}{s^2R_x(C_x+C_p+C_i)}\Bigg( \frac{1+sC_xR_x}{s+\Big(\dfrac{A_o \omega_0 R_x C_i +1}{R_x(C_x+C_p+C_i)}\Big)}\Bigg)
    \label{Eq_modified1}
\end{align}
\begin{figure*}[!h]
\small

\begin{equation}
\begin{split}
    V_R(t) =&  V_p(\alpha - 2X) - \frac{V_p}{R_x C_i} \Bigg\{ \overbrace{\frac{t}{\Big(1+\dfrac{1}{A_0 \omega_0R_x C_i}\Big)}}^\text{{Integration slope reduction due to finite GBW}} - \\
    & \underbrace{\frac{1}{\Big(1+\frac{1}{A_0 \omega_0R_x C_i}\Big)^2} \frac{1}{A_0\omega_0} \Big[ 1+\big(\frac{C_x+C_p}{C_i}\big)+\frac{C_x}{C_i}(1+A_0 \omega_0R_x C_i) \Big]e^{-\Big(\tfrac{A_o \omega_0 R_x C_i +1}{R_x(C_x+C_p+C_i)} \Big)t}}_\text{{Offset voltage due to finite GBW}}\Bigg\}
\end{split}
\label{eq:time_domain_1}
\end{equation}
\hrulefill
\end{figure*}
The time domain voltage $\mathrm{V_R(t)}$ is obtained by transforming Eq.~\ref{Eq_modified1} into the time-domain and taking the initial value of $\mathrm{V_R(t)}$ as $\mathrm{V_P(\alpha - 2X)}$ where, $\mathrm{\alpha = R_2/R_1}$ and $X=C_x/C_i$. {The time-domain voltage expression at node $V_R$ is derived in Eq. \ref{eq:time_domain_1}.} The integrator output voltage contains additional terms due to the Op-Amp non-idealities as seen from Eq.~\ref{eq:time_domain_1}. 
\begin{itemize}
    \item The first term indicates that the finite GBW product reduces the slope of the linear ramp voltage at the integrator output by a factor $\mathrm{1+1/(A_o \omega_0 R_x C_i)}$.
    \item  The second term offers insight into the offset voltage introduced at the integrator output due to the finite GBW product and additionally due to the impedance/leaky-capacitive  sensor. The effect of the exponential transient term is small enough to be neglected. We denote the final offset term introduced due to finite GBW by $\gamma$ and is given by
    \end{itemize}
\begin{equation}
        \gamma = \Bigg(1+\frac{C_x+C_p}{C_i}+\frac{C_x}{C_i}(1+A_0\omega_0R_xC_i)  \Bigg)\frac{A_0\omega_0R_xC_i}{(1+A_0\omega_0R_xC_i)^2}
    \label{Eq_gamma}
    \end{equation}

Further, considering the Op-Amp in the integrator has an offset voltage of $V_{os}$ and a bias current of $i_b$, the total offset voltage presented by the  integrator is then $\mathrm{V_{os}' = V_{os} + i_bR_x}$. 
A similar analysis of the integrator with a finite GBW Op-Amp in the presence of this offset term shows that the effect is evaluated simply by shifting the input by the equivalent offset voltage. Further, the slope of the integrator output waveform (triangular) is limited by the slew rate (SR) of the Op-Amp used in the integrator. Hence, the slope (SL) of the integrator output is given by Eq.~\ref{Eq_SL}, where the positive sign corresponds to $SL_{(-)}$, and the minus sign corresponds to $SL_{(+)}$.
\begin{align}
    SL = min \Bigg\{ \frac{V_p \pm (V_{os}'+i_bR_x)}{R_xC_i \Big( 1+\dfrac{1}{A_0\omega_0R_xC_i}  \Big)}, SR \Bigg\} 
    \label{Eq_SL}
\end{align}

Referring to Fig.~\ref{fig:circuit_5}, we now obtain the individual expressions for $T_{P1}$ and $T_{P2}$ to obtain the $R_x$ and $C_x$ value, which is the measurement parameter. Referring to Fig.~\ref{fig:circuit_5}, $T_{P1}$ {can be expressed as},
\begin{align}
\begin{split}
    T_{P1} = max\Bigg\{\Bigg[ \frac{(\alpha - \gamma - 2X - V_{oz}/V_p)}{1 + (V_{os}'+i_bR_x)/V_p}\Bigg] 
    R_xC_i \\ \cdot \Bigg(1+\dfrac{1}{A_0\omega_0R_xC_i}\Bigg), 
    \frac{(\alpha-2X) V_p}{SR} \Bigg\}
    \label{EQ_TP1_1}
\end{split}
\end{align}
Similarly,  $T_{P2}$ can be obtained as,
\begin{align}
\begin{split}
    T_{P2} = max\Bigg\{\Bigg[ \frac{(\alpha + \gamma + V_{oz}/V_p)}{1 + (V_{os}'+i_bR_x)/V_p}\Bigg] R_xC_i \\ \cdot \Big(1+\dfrac{1}{A_0\omega_0R_xC_i}\big),  \frac{(\alpha-2X) V_p}{SR} \Bigg\}
    \label{EQ_TP2_1}
\end{split}
\end{align}

\begin{figure*}[!b]
\small
\begin{align}
    T_{P1} = max\Bigg\{ R_xC_i(\alpha-2X)\Bigg[\frac{1-\dfrac{\gamma+V_{oz}/V_p}{\alpha-2X}}{(1+V_{os}'/V_p)} \Bigg]\Big(1+\dfrac{1}{A_0\omega_0R_xC_i}\Big), \frac{(\alpha-2X) V_p}{SR}
    \Bigg\} + \frac{\tau_{S,LH}+\tau_{Z,LH}}{2}
    \label{EQ_TP1_2}
\end{align}
\end{figure*}

\begin{figure*}[!t]
\small
\begin{align}
    T_{P2} = max\Bigg\{ R_xC_i\alpha\Bigg[\frac{1+\dfrac{\gamma+V_{oz}/V_p}{\alpha}}{(1+V_{os}'/V_p)} \Bigg] \Big(1+\dfrac{1}{A_0\omega_0R_xC_i}\Big), \frac{(\alpha-2X) V_p}{SR}
    \Bigg\} + \frac{\tau_{S,HL}+\tau_{Z,LH}}{2}
    \label{EQ_TP2_2}
\end{align}
\hrulefill
\end{figure*}

\textbf{ Timing non-idealities in the ZCD and the Schmitt trigger}
Considering the rise and fall times of the Schmitt trigger and the ZCD as: $\tau_{S,HL}$ and $\tau_{S,LH}$ be the high to low and low to high response delays of the Schmitt trigger and $\tau_{Z,HL}$ and $\tau_{Z,LH}$ be high to low and low to high response delays of the ZCD, the expression of $T_{P1}$ and $T_{P2}$ in Eq.~\ref{EQ_TP1_1} and Eq.~\ref{EQ_TP2_1}, are refined as Eq.~\ref{EQ_TP1_2} and Eq.~\ref{EQ_TP2_2}, respectively. 

\textbf{Combined effect of the non-idealities on the measurement}
From the combined non-idealities represented in Eq.~\ref{EQ_TP2_2}, the expression for the sensor resistance $R_x$ can be inferred as follows.
\begin{align}
    R_x =  \frac{T_{P2}}{\alpha C_i} \Bigg[ \frac{1}{\beta_1}-\frac{\alpha}{A_0\omega_0}\Bigg]
    \label{EQ_RX_TP2}
\end{align}
where 
\begin{equation}
    \beta_1 = \frac{1+(\gamma+V_{oz}/V_p)/\alpha}{1+V_{os}'/V_p}
    \label{Eq_beta1}
\end{equation}
The expression for the $R_x$ shows that the sensor parameter is significantly affected by the non-idealities of the circuit components. Similarly, for the capacitive sensors and parallel R-C sensors, the measurement capacitor $C_x$ can be derived from Eq. \ref{EQ_TP1_2} and Eq.~\ref{EQ_TP2_2} as follows.
\begin{equation}
\begin{split}
    C_x = \frac{\alpha C_i}{2+\Bigg(\dfrac{A_0\omega_0R_xC_i}{1+A_0\omega_0R_xC_i}\Bigg)}  \Bigg[& 1-\Big(\frac{T_{P1}}{T_{P2}}\Big)\Bigg(\frac{1+V_{os}'/V_p}{\dfrac{1}{\beta_1}-\dfrac{\alpha}{A_0\omega_0}}\Bigg) \\ & \Big(\dfrac{1}{1+\dfrac{1}{A_0\omega_0R_xC_i}}\Big) -\frac{V_{oz}}{\alpha V_p} \Bigg]
\end{split}
\label{EQ_CX_TP1}
\end{equation}

Eq.~\ref{EQ_RX_TP2} and Eq.~\ref{EQ_CX_TP1} provide the measured sensor capacitance and sensor resistance values $R_x$ and $C_x$ with the component non-idealities. The ideal values of measured sensor resistance and capacitance, are: $\mathit{R_x} = \dfrac{T_{P2}}{\alpha C_i}$ and, $\mathrm{C_x} = \dfrac{\alpha C_i}{2}(1-\dfrac{T_{P1}}{T_{P2}})$. Comparing the $\mathrm{R_x}$ and $\mathrm{C_x}$ values obtained from Eq.~\ref{EQ_RX_TP2} and Eq.~\ref{EQ_CX_TP1} with the ideal values, it is evident that there is a large error introduced due to the component non-idealities.

The analytical solution for $C_x$ and $R_x$ derived in Eq.~\ref{EQ_CX_TP1}  and Eq.~\ref{EQ_RX_TP2} is evaluated by including the non-idealities of commercial operational amplifiers. The results show that even for a precision op-amp such as OPA177FP from Analog Devices, the worst-case error in the measurement due to component non-idealities is as high as 30\% for $R_x$ measurement
and 60\% for $C_x$ measurement.

\textbf{Accuracy Enhancement with Single Cycle Averaging}\\
\begin{figure*}[b]
\hrulefill
\small
\begin{align}
    T_{P3} = max\Bigg\{ R_xC_i(\alpha-2X)\Bigg[\frac{1+\dfrac{\gamma+V_{oz}/V_p}{\alpha-2X}}{(1-V_{os}'/V_p)} \Bigg] \Bigg(1+\dfrac{1}{A_0\omega_0R_xC_i}\Bigg), \frac{(\alpha-2X) V_p}{SR}
    \Bigg\} + \frac{\tau_{S,HL}+\tau_{Z,HL}}{2}
    \label{EQ_TP3_2}
\end{align}
\end{figure*}

\begin{figure*}[t]
\small
\begin{align}
    T_{P4} = max\Bigg\{ R_xC_i\alpha\Bigg[\frac{1-\dfrac{\gamma+V_{oz}/V_p}{\alpha}}{(1-V_{os}'/V_p)} \Bigg] \Bigg(1+\dfrac{1}{A_0\omega_0R_xC_i}\Bigg), \frac{(\alpha-2X) V_p}{SR}
    \Bigg\} + \frac{\tau_{Z,HL}+\tau_{S,LH}}{2}
    \label{EQ_TP4_2}
\end{align}
\hrulefill
\small
\end{figure*}
\begin{figure}[ht]
    \centering
    \includegraphics[trim=15 18 15 18, clip, scale = 1.2]{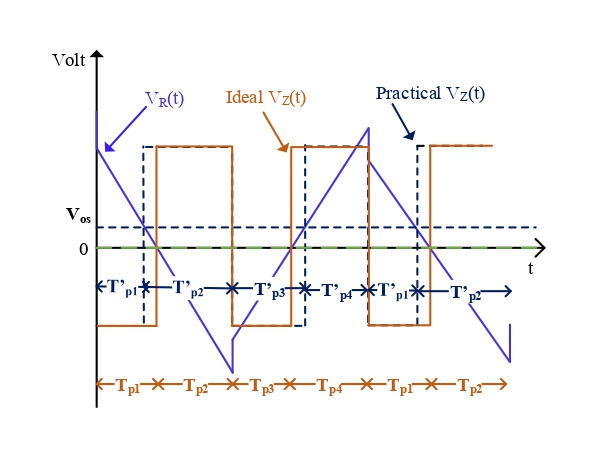}
     \caption{  The ideal and practical $V_z(t)$ of the proposed relaxation oscillator based circuit shown in Fig. \ref{fig:circuit_4}. The offset due to the component non-idealities is represented as $V_{os}$. This offset affects the period $T_{p1}$, $T_{p2}$, $T_{p3}$, and $T_{p4}$.The practical periods due the offset is represented by $T'_{p1}$, $T'_{p2}$, $T'_{p3}$, and $T'_{p4}$. Averaging the alternate cycles of practical $V_z(t)$ significantly reduces the error.
}   
     \label{fig:ideal_prac}
\end{figure}

The timing waveform of the proposed circuit with an offset voltage $V_{os}$ due to component non-idealities is shown in Fig. \ref{fig:ideal_prac}. The ideal timing waveform $T_{p1}$, $T_{p2}$, $T_{p3}$, and $T_{p4}$ are shown in Fig. \ref{fig:ideal_prac} as ideal $V_z(t)$. However, as derived in Eq. \ref{eq:time_domain_1}, the zero-crossing point of $V_R(t)$ is shifted by an offset voltage $V_{os}$ due to component non-idealities such as fine GBW and offset voltage of operational amplifiers. This affects the practical $V_z(t)$ as shown in Fig. \ref{fig:ideal_prac}. Consider $T_{of}$ as the corresponding offset time between the ideal and the practical output waveform $V_z(t)$ due to finite offset voltage $V_{os}$. From Fig. \ref{fig:ideal_prac}, if $T_{p1}$ switches from $-V_p$ to $+V_p$ by $T_{of}$ time earlier than ideal, then $T_{p3}$ will switch from $+V_p$ to $-V_p$ after $T_{of}$ time compared to the ideal transition.

Therefore, the practical expression for time periods can be written as $T'_{p1} = T_{p1} - T_{of}$ and $T'_{p3} = T_{p3}+T_{of}$. The averaging of $T'_{p1}$ and $T'_{p3}$ will eliminate the effect of the mismatch in the time period between $T_{p1}$ and $T_{p3}$. Same is true for $T_{p2}$ and $T_{p4}$. Therefore, intuitively, it is possible to reduce the effect of component non-idealities by single-cycle averaging. It can be seen from the waveform that the pattern repeats after $T'_{p4}$. Therefore, averaging more than one cycle does not further enhance the accuracy of the sensor parameter measurement due to component non-idealities.

In order to analyze the effect of averaging, we have derived the expression for the period $T_{p3}$ and $T_{p4}$ in Eq. \ref{EQ_TP3_2} and Eq. \ref{EQ_TP4_2}, respectively. The measurement of $T_{p3}$ and $T_{p4}$ is useful to understand the effect of non-idealities in the subsequent cycle of one complete cycle of $V_x(t)$.

The time periods $\mathrm{T_{p1}}$ and $\mathrm{T_{p3}}$ are averaged to obtain $\mathrm{T_1}$ and, the time periods $\mathrm{T_{p2}}$ and $\mathrm{T_{p4}}$ are averaged to obtain $\mathrm{T_2}$. 
The expression of the averaged period $T_1$ and $T_2$ are derived in Eq. \ref{EQ_T1_1} and Eq. \ref{EQ_T2_1},({on page 11}) respectively. The high-to-low and low-to-high response delays are assumed to be the same for the ZCD ($\mathrm{\tau_Z}$) and Schmitt trigger ($\mathrm{\tau_S}$) for deriving Eq.~\ref{EQ_T1_1} and Eq.~\ref{EQ_T2_1}. 

\begin{figure*}[!t]
\small
\hrulefill
\begin{equation}
\begin{split}
    T_{1} = \frac{T_{P1}+T_{P3}}{2} = max\Bigg\{ R_xC_i(\alpha-2X)\Bigg[\frac{1+(\dfrac{V_{os}'}{V_p})\big(\dfrac{\gamma+V_{oz}/V_p}{\alpha-2X}\big)}{1-(V_{os}'/V_p)^2} \Bigg] \Bigg(1+\dfrac{1}{A_0\omega_0R_xC_i}\Bigg), \frac{2V_p(\alpha-X)}{SR}
    \Bigg\} + \frac{\tau_{S}+\tau_{Z}}{2}
    \label{EQ_T1_1}
\end{split}
\end{equation}
\end{figure*}

\begin{figure*}[!t]
\small
\begin{equation}
\begin{split}
    T_{2} = \frac{T_{P2}+T_{P4}}{2} = max\Bigg\{ \alpha R_xC_i\Bigg[\frac{1-(\dfrac{V_{os}'}{\alpha V_p})\big(\gamma+V_{oz}/V_p\big)}{1-(V_{os}'/V_p)^2} \Bigg] \Bigg(1+\dfrac{1}{A_0\omega_0R_xC_i}\Bigg), \frac{2V_p(\alpha-X)}{SR}
    \Bigg\} + \frac{\tau_{S}+\tau_{Z}}{2}
    \label{EQ_T2_1}
\end{split}
\end{equation}
\hrulefill
\end{figure*}


To observe and compare the enhancement in the accuracy due to averaging, consider the case when all the offsets are zero, the SR is not limited, and the only non-ideality is the limit on the GBW. For this case, the time expressions without and with averaging are derived in the subsequent subsections.

\textbf{{Without Averaging: $T_{p1}$ and $T_{p2}$}}
The Eq. \ref{EQ_TP1_GBW} and Eq. \ref{EQ_TP2_GBW} ({on page 12}) shows that the period $T_{p1}$ and $T_{p2}$ for the measurement of the sensor parameters are greatly affected by the gain-bandwidth products of the operational amplifiers. For instance, for OPA177, for an R$_x$ of 330 k$\Omega$ and C$_x$ of 33 pF, the error in the period $T_{p1}$ and $T_{p2}$ due to the GBW is 52.76$\%$ and 26.74$\%$, respectively.

\begin{align}
    T_{P1} &= R_xC_i(\alpha-2X) \Bigg(1-\dfrac{\gamma}{\alpha-2X}\Bigg) \Bigg(1+\dfrac{1}{A_0\omega_0R_xC_i} \Bigg)\label{EQ_TP1_GBW}\\ 
    T_{P2} &= \alpha R_xC_i \Bigg(1+\dfrac{\gamma}{\alpha}\Bigg) \Bigg(1+\dfrac{1}{A_0\omega_0R_xC_i} \Bigg) \label{EQ_TP2_GBW}
\end{align}

\textbf{With Single Cycle Averaging: $T_1$ and $T_2$}
Considering only the GBW of the operational amplifiers, $T_1$ and $T_2$ from Eq. \ref{EQ_T1_1} and Eq. \ref{EQ_T2_1} can be simplified as follows. 
\begin{align}
    T_{1} &= R_xC_i(\alpha-2X) \Bigg(1+\dfrac{1}{A_0\omega_0R_xC_i} \Bigg) \label{EQ_T1_GBW}\\ 
    T_{2} &= \alpha R_xC_i \Bigg(1+\dfrac{1}{A_0\omega_0R_xC_i} \Bigg) \label{EQ_T2_GBW}
\end{align}
The offset, $\mathrm{\gamma}$ introduced due to the finite GBW of the Op-Amp severely affects the accuracy of time measurement as observed in  Eq.~\ref{EQ_T1_GBW} and Eq.~\ref{EQ_T2_GBW} along with the slope reduction factor of $\mathrm{1+1/(A_o \omega_0 R_x C_i)}$. With averaging, the time accuracy is not dependant on the offset factor $\mathrm{\gamma}$ and depends only on the slope reduction factor as observed in Eq.~\ref{EQ_T1_GBW} and Eq.~\ref{EQ_T2_GBW}. Therefore, the single-cycle averaging significantly reduces the error due to the GBW of the Op-Amps. For example, in the case of OPA177, with an R$_x$ of 330 k$\Omega$ and C$_x$ of 33 pF, the error in the period $T_{1}$ and $T_{2}$ due to the GBW is 1.5$\%$ and 0.72$\%$, respectively, which  is significantly smaller as compared to the measurement without averaging.

\textbf{Design Criteria}\\
The first design criterion is the selection of $\mathrm{\alpha}$, which decides the range of measurement. A higher value of $\mathrm{C_x}$ results in a higher factor $X$ (i.e. $C_x/C_i$), which might result in a cross-over of the voltage beyond the zero value while charge transfer. To avoid this, we get the following condition on $\mathrm{\alpha}$
\begin{equation}
    \alpha > 2\frac{C_{x,max}}{C_i}
\end{equation}

The error in capacitance measurement $\mathrm{\epsilon}$ results in a worst-case error compared to the error in the resistance measurement. The error in capacitance measurement is the deviation of time $\mathrm{T_1}$ from the ideal value. 
\begin{align}
    \epsilon = \frac{\Delta C_x}{C_x} = \frac{\Delta T_1}{T_1}
    \label{Eq_delt_by_t}
\end{align}
{$\Delta C_x$ represents the incremental value of sensor capacitance and $\Delta$ $T_1$ represents the corresponding relative change measurement time}. \\
Using Equations~\ref{EQ_T1_1},~\ref{Eq_gamma} and,~\ref{Eq_delt_by_t} the following conditions on the Op-Amp GBW, SR, offset can be arrived at,

 \begin{equation}
       \footnotesize{ \epsilon < \frac{100}{1-\Bigg(\dfrac{V_{os}'}{V_p}\Bigg)^2} \Bigg[ \frac{1}{GBW(R_{x,min}C_i)}+ \\
      \frac{V_{os}'}{V_p}\Bigg(\frac{1}{\dfrac{\alpha C_i}{C_{x,max}}-2 } + \frac{V_{os}'}{V_p} \Bigg)   \Bigg] \label{Eq_req_opamp}}
\end{equation}

\begin{align}
    SR  &> \frac{2 V_p}{R_{x,min}C_i} 
    \label{Eq_req_SR}
\end{align}

For a given error tolerance in the measurement, an Op-Amp with a GBW and offset requirement and SR that satisfies Eq.~\ref{Eq_req_opamp} and Eq.~\ref{Eq_req_SR} must be chosen. A comparator must also be selected with an offset value $\mathrm{V_{oz} << V_p}$ and response time ($\mathrm{\tau}$) condition derived as follows. 
\begin{align}
    \tau < \frac{\epsilon \alpha R_{x,max} C_i}{100\Big( 1- (V_{os}'/V_{p})^2\Big)}
    \label{eq:req_tau}
\end{align}

Considering the propagation time of the XOR gate as $\tau_p$, the XOR gate must be chosen based on the following equation.

\begin{align}
    \tau_p < \frac{\epsilon \alpha R_{x,max} C_i}{100\Big( 1- (V_{os}'/V_{p})^2\Big)} + \tau
    \label{eq:req_tau_p}
\end{align}

\textbf{Experimental Setup and Results}\\
Several prototypes of the proposed relaxation oscillator-based signal conditioning circuit shown in Fig. \ref{fig:circuit_4} are built to illustrate the circuit's capabilities for measuring sensor capacitance and resistance. Different operational amplifiers are employed to implement the integrator, Schmitt trigger, and ZCD of the proposed oscillator and are tabulated in Table \ref{Table:components}, further using these opamps, the effect of component non-idealities on the measurement of sensor parameters has been studied. All the passive components are first measured using a commercial table-top LCR meter {(Agilent E4980A)}.
\begin{table}[t!]
\centering 
\caption{{Operational amplifiers with performance parameters}}
{\begin{tabular}{|ccccccc|}
\hline
\textbf{\begin{tabular}[c]{@{}c@{}}Op-Amp \\ IC\end{tabular}} & \textbf{\begin{tabular}[c]{@{}c@{}}GBW/(2$\pi$)\\ (MHz)\end{tabular}} & \textbf{\begin{tabular}[c]{@{}c@{}}SR\\ ($V/\mu s$)\end{tabular}} & \textbf{\begin{tabular}[c]{@{}c@{}}$C_p$\\ (pF)\end{tabular}} & \textbf{\begin{tabular}[c]{@{}c@{}}$V{of}$\\ (mV)\end{tabular}} & \textbf{\begin{tabular}[c]{@{}c@{}}$i_b$\\ (nA)\end{tabular}} & \textbf{\begin{tabular}[c]{@{}c@{}}$\tau$\\ (ns)\end{tabular}} \\ \hline
\textbf{AD741}                                                  & 1 & 0.5  & --  & 5    & 500    & 300   \\ \hline
\textbf{LT1360} & 60 & 800   & 4  & 0.3  & 250 & --  \\ \hline
\textbf{TL071} & 5.25  & 29 & 2 & 4  & 0.1 & 310 \\ \hline
\textbf{OPA177} & 0.6 & 0.3 & -- & 0.6 & 6 & -- \\ \hline
\textbf{LT1049} & 0.8 & 0.8 & -- & 0.01 & 0.05 & -- \\ \hline
\end{tabular}}
\label{Table:components}
\end{table}
The periods of the readout signal are measured using the timer/counter module of the microcontroller. In this experiment, we have used an Arduino Uno microcontroller with a 16-bit timer. The data from the microcontroller is acquired in MATLAB, and the single-cycle averaging is performed. 

Multiple experiments were conducted to evaluate the performance of the proposed signal conditioning circuit. {Initial} experiments were conducted by emulating the sensor using discrete resistors and capacitors. The final experiment is conducted on a fringing field-based capacitive water level sensor.  
The sensor is emulated by placing known values of resistor $R_x$ and capacitor $C_x$ in parallel. The time period $T_{p1}$, $T_{p2}$, $T_{p3}$, and $T_{p4}$ are measured to calculate the value of the sensor resistor. \\
\textbf{Resistance Measurement}

\begin{figure*}[b]
    \centering
    \includegraphics[width=140mm, height=70mm]{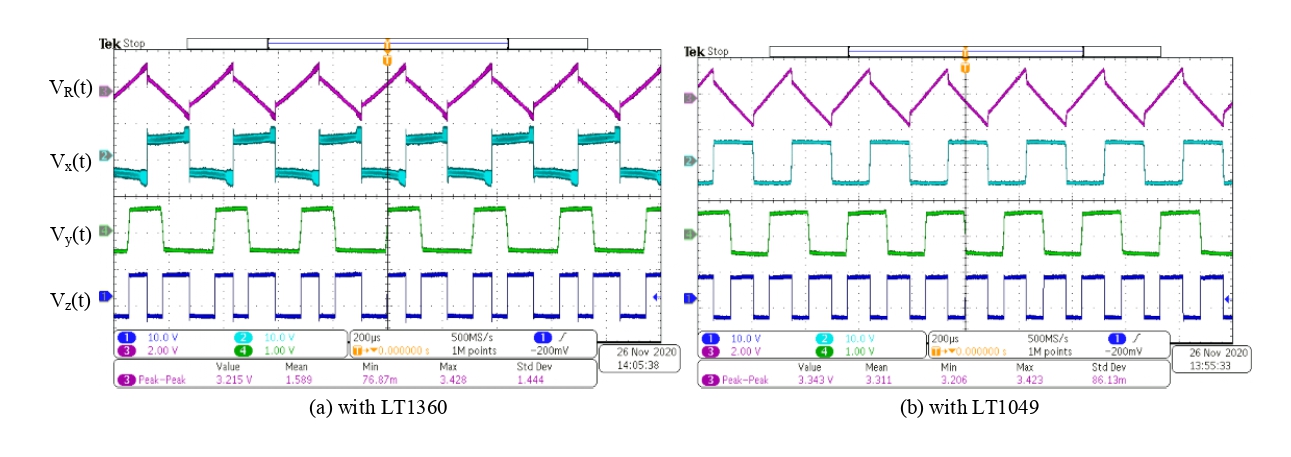}
     \caption{The waveform at different nodes of the oscillator developed using op-amp LT1360 and LTC1049. The voltage $V_R(t)$ is affected by the non-idealities of the op-amps, which affects the pulse width of $V_Y(t)$ (ideally should be 50\% duty cycle). Due to this, the output $V_z(t)$ has different positive and negative pulse widths resulting in an error in the measurement of sensor parameters. (a) The error in the measurement is significant for LT1360, (b) the error is reduced with LTC1049 compared to LT1360. The error is negligible after a single-cycle averaging operation.}   
     \label{fig:circuit_7}
\end{figure*}

\begin{figure}
    \centering
    \includegraphics[trim=0 0 0 0, scale = 0.1]{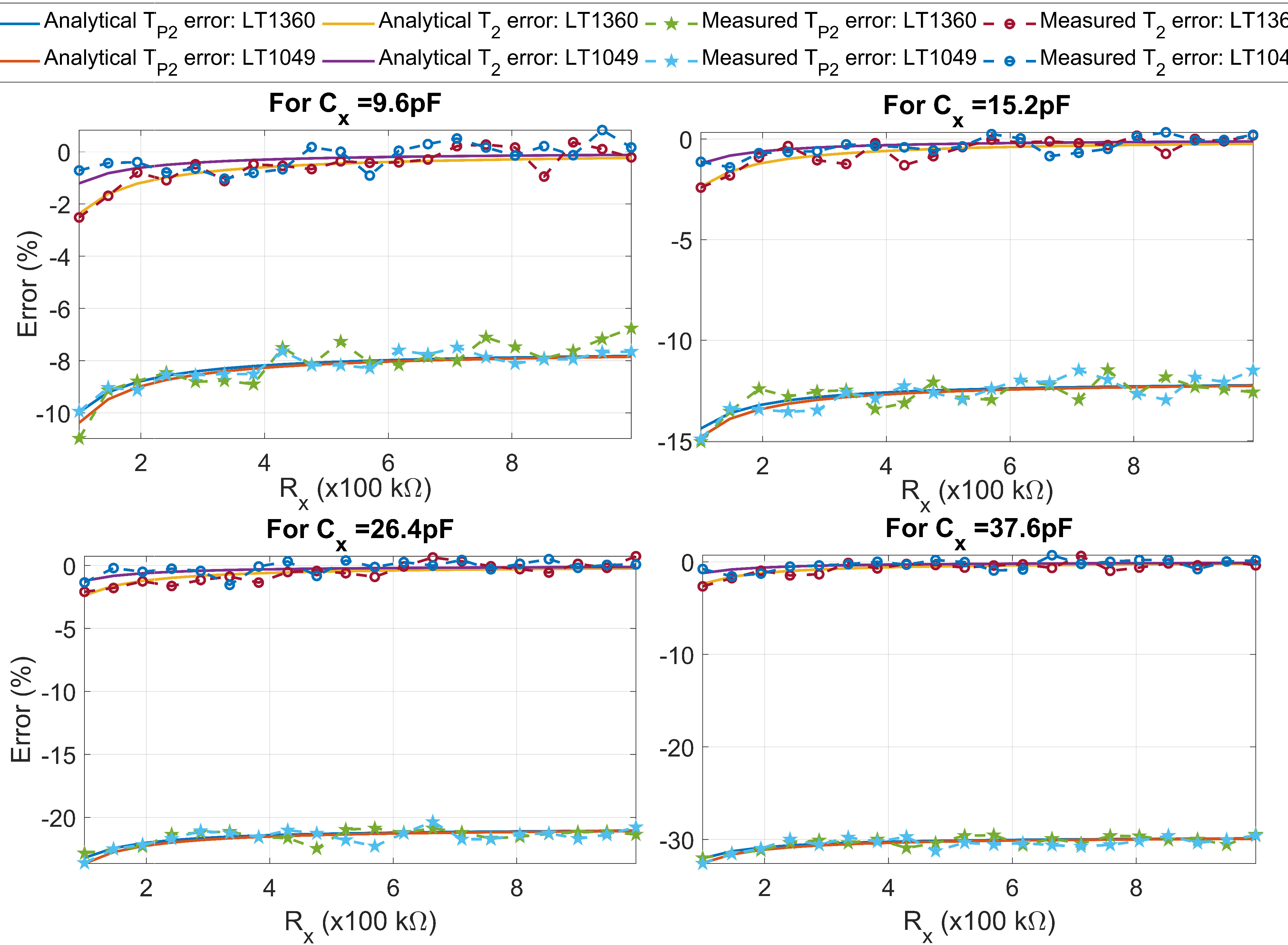}
          \caption{Measured and Analytical error in time ($T_2 \& T_{P2}$) for different sensor capacitance $C_x$ and resistance $R_x$ for two the LT1360 and LTC1049. The analytical $T_{p2}$ error is calculated from Eq. \ref{EQ_TP2_2}, which shows the error in the measurement due to component non-idealities. The analytical $T_2$ error shows the effect of single cycle averaging and is calculated from Eq. \ref{EQ_T2_1}. The analytical and experimental error reduces significantly with single-cycle averaging.}   
     \label{fig:circuit_9}
\end{figure}

\begin{figure}[ht]
    \centering
    \includegraphics[trim=0 0 0 0, scale = 0.1]{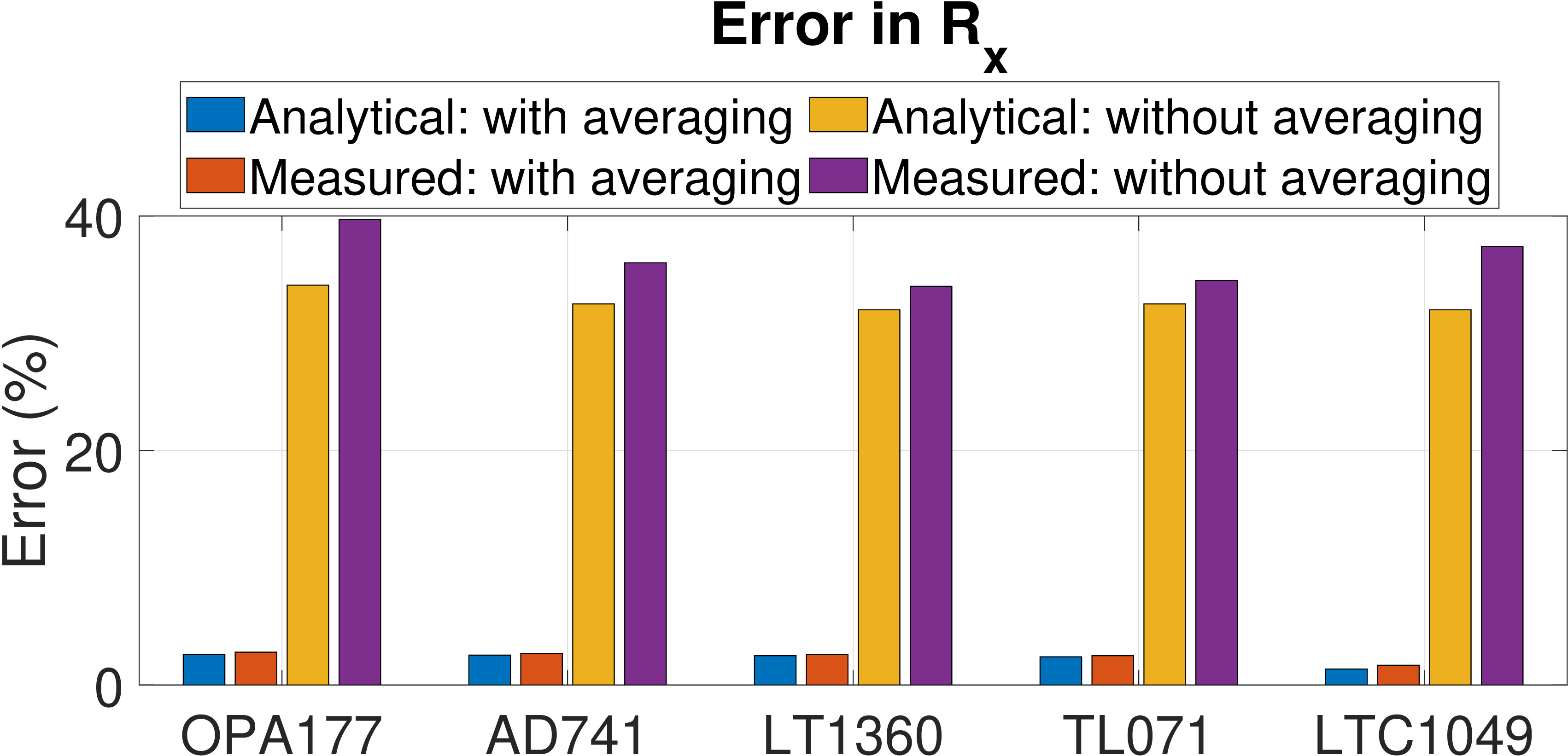}
     \caption{Analytical and measured worst-case error in sensor resistance $R_x$ with different Op-Amps. The analytical error without averaging is calculated from Eq. \ref{EQ_TP2_2}. The analytical error with averaging is shown the effect of single cycle averaging and calculated from Eq. \ref{EQ_T2_1}.}   
     \label{fig:circuit_9b}
\end{figure}
\begin{figure}
    \centering
    \includegraphics[trim=0 0 0 0, scale = 0.1]{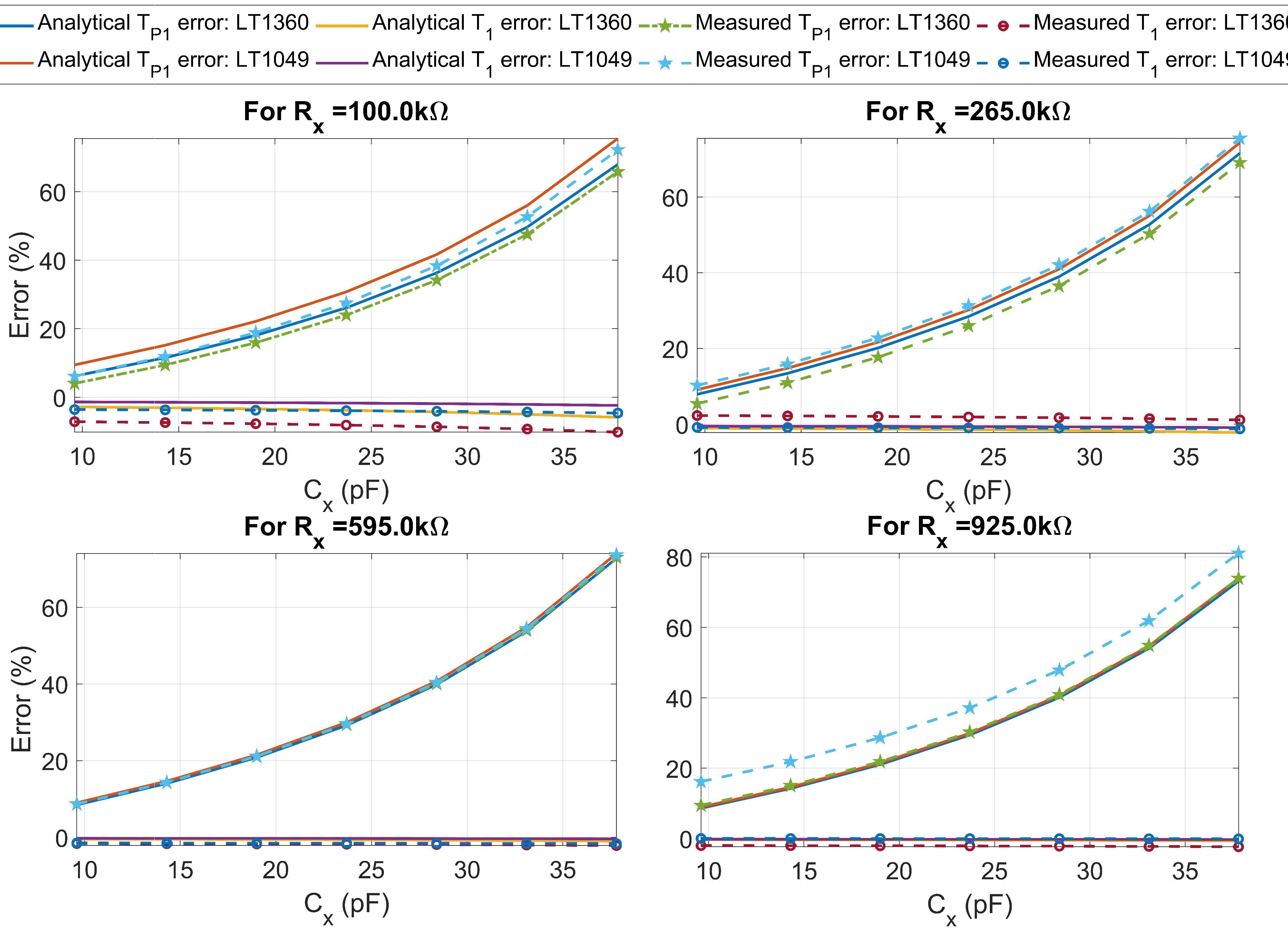}
     \caption{Measured and Analytical error for different sensor capacitance $C_x$ and resistance $R_x$ for two different Op-Amps.Analytical and measured worst-case error in sensor capacitance $C_x$ with different Op-Amps. The analytical $T_{p1}$ and $T_1$ errors are calculated from Eq. \ref{EQ_TP1_1} and Eq. \ref{EQ_T1_1}, respectively. The analytical $T_{p1}$ shows the error due to component non-idealities. The analytical $T_2$ error shows the effect of single cycle averaging. The results are experimentally verified.}   
    \label{fig:circuit_8}
\end{figure}
The resistance ($R_x$) measurement is performed by measuring the period $T_{p2}$ and $T_2$ (average of $T_{p2}$ and $T_{p4}$). In addition to the measurement of the sensor resistance, the experiment also demonstrates the effect of component non-idealities on the measurement. The experiment is conducted as follows. 

\begin{enumerate}
    \item A capacitor $C_x$ is placed in parallel with the sensor resistance $R_x$. The value of $R_x$ is varied from 100 k$\Omega$ to 1000 k$\Omega$. The period $T_{p2}$ and $T_{p4}$ is measured, and the percentage relative error is calculated with and without averaging.
    \item The same process is repeated for multiple values of the capacitance $C_x$. 
    \item The above process is repeated for different operational amplifiers, and the effect of component non-idealities are evaluated on the measurement. 
\end{enumerate}

The percentage relative error for different values of the sensor resistance at different $C_x$ values for the operational amplifiers LT1360 and LTC1049 are shown in Fig. \ref{fig:circuit_9}. The result shows that the single cycle averaging significantly reduces the error in the measurement of the sensor resistance. The worst-case percentage relative error with and without single cycle averaging for R$_x$ measurement, for different operational amplifiers is shown in Fig. \ref{fig:circuit_9b}. The chart in Fig. \ref{fig:circuit_9b} shows the choice of the operational amplifier and the effect of non-idealities. The \% relative error is high for the operational amplifier OPA177, which has a low gain-bandwidth. Moreover, the \% relative error is also high for LT1360, which has a good gain-bandwidth but has high input offset voltage. On the other hand, the error is small for LTC1049, which has a moderate gain-bandwidth and small input offset voltage. Therefore, the oscillator should be designed considering all the non-idealities of the operational amplifiers. The operational amplifiers' selection should be done based on the Eq. (\ref{Eq_req_SR}) and Eq. (\ref{eq:req_tau_p}) from the design criteria for the given error.\\

\begin{table}
\caption{Performance Parameters of the Developed Circuit}
\begin{tabular}{|cccc|}
\hline
\multicolumn{1}{|c|}{\multirow{2}{*}{\textbf{Parameters}}}                                     & \multicolumn{1}{c|}{\multirow{2}{*}{\textbf{Expression}}}                                         & \multicolumn{2}{c|}{\textbf{Measurement}}                \\ \cline{3-4} 
\multicolumn{1}{|c|}{}                                                                         & \multicolumn{1}{c|}{}                                                                             & \multicolumn{1}{c|}{$C_x$}     & $R_x$                   \\ \hline
\multicolumn{1}{|c|}{\textbf{SD($\sigma$)}}                                                    & \multicolumn{1}{c|}{$\sqrt{\sum_{n=1}^{M} \frac{(S(n)-\overline{S})^2}{M-1}}$}                    & \multicolumn{1}{c|}{10.2fF}    & 8.1 $\Omega$            \\ \hline
\multicolumn{1}{|c|}{\textbf{SNR}}                                                             & \multicolumn{1}{c|}{$10\log\frac{\sum_{n=1}^{M} (S(n))^2}{\sum_{n=1}^{M} (S(n)-\overline{S})^2}$} & \multicolumn{1}{c|}{74.83 db}  & 78.96 db                \\ \hline
\multicolumn{1}{|c|}{\textbf{Dynamic Range}}                                                   & \multicolumn{1}{c|}{-}                                                                            & \multicolumn{1}{c|}{10pF-42pF} & 100k$\Omega$-1M$\Omega$ \\ \hline
\multicolumn{1}{|c|}{\textbf{\begin{tabular}[c]{@{}c@{}}Worst \\ Relative Error\end{tabular}}} & \multicolumn{1}{c|}{-}                                                                            & \multicolumn{1}{c|}{1.6 $\%$}  & 0.62$\%$                \\ \hline
\multicolumn{4}{|c|}{$S(n)$ is the $n^{th}$ measured value $\bar{S}$ is the average value of the measurements}                                                                                                                \\ \hline
\end{tabular}
\label{table:performace_parameters}
\end{table}

\textbf{Capacitance Measurement}\\
In case of capacitance measurement, the value of sensor capacitor $C_x$ is varied and the periods $T_{p1}$, $T_{p2}$, $T_{p3}$, and $T_{p4}$ are measured experimentally. The experiment is conducted for different values of $R_x$. The experimental procedure for $C_x$ measurement of impedance R-C or leaky capacitive sensors is as follows.
\begin{enumerate}
    \item  First, $R_x$ is fixed at one value and $C_x$ is varied from 9.6 pF to 37.6 pF.
    \item   Next, the same process is repeated for different $R_x$ values. The experiment is conducted using different operational amplifiers. 
    \item The experimental results for different values of sensor capacitor $C_x$ for operational amplifiers LT1360 and LTC 1049 are shown in Fig.~\ref{fig:circuit_8}.
\end{enumerate}

\begin{figure}
    \centering
    \includegraphics[trim=0 0 0 0, scale = 0.1]{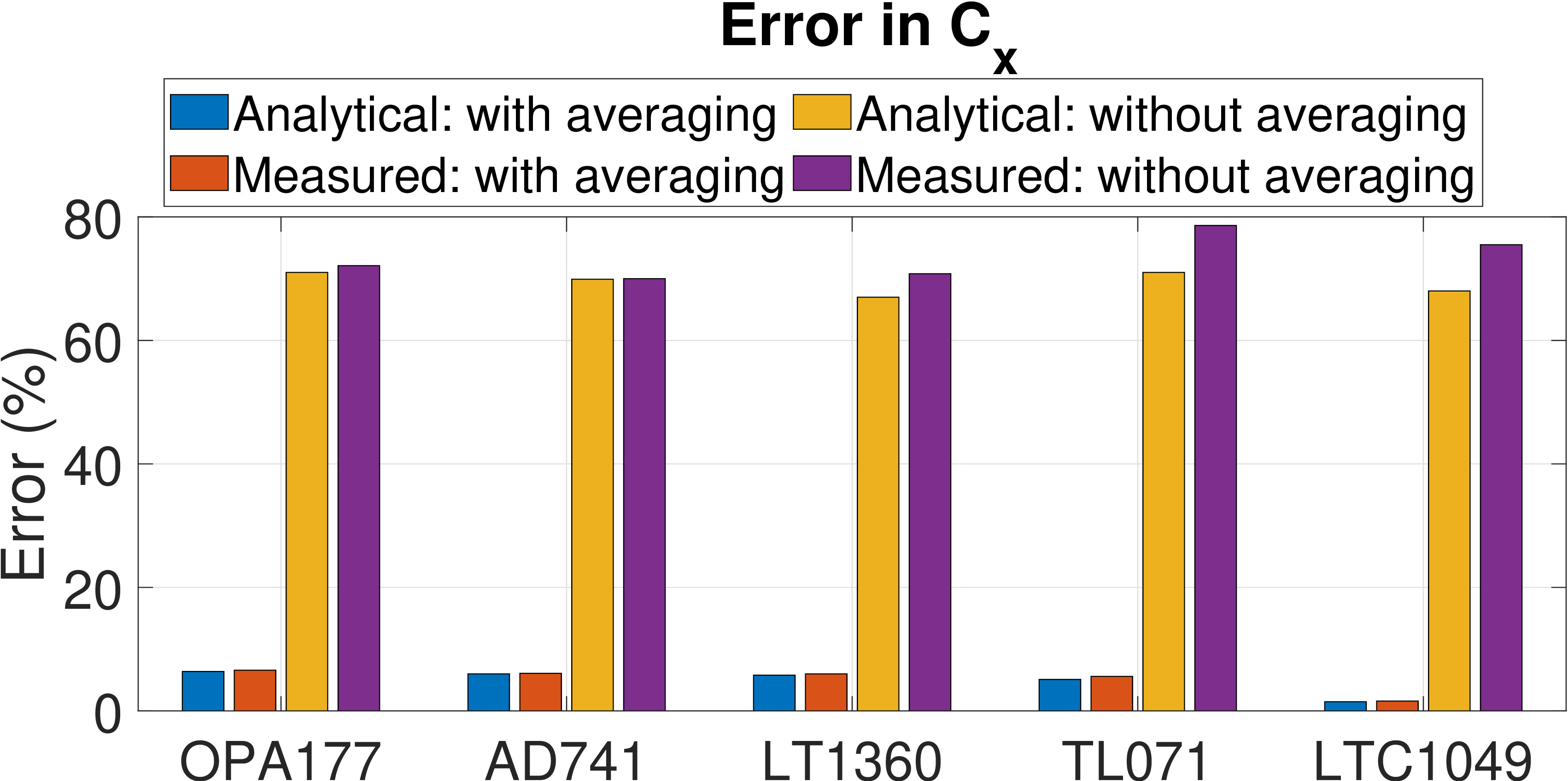}
     \caption{Analytical and measured  worst-case error in sensor capacitance $C_x$ with different Op-Amps. The analytical $T_{p1}$ and $T_1$ errors are calculated from Eq. \ref{EQ_TP1_1} and \ref{EQ_T1_1}, respectively. The analytical $T_{p1}$ shows the error due to component non-idealities. The analytical $T_2$ error shows the effect of single cycle averaging. The results are experimentally verified.}   
     \label{fig:circuit_9a}
\end{figure}

 The error is calculated from the measured time periods with and without single cycle averaging. The worst-case errors for different operational amplifiers are shown in Fig. \ref{fig:circuit_9a}. The single-cycle averaging dramatically reduces the percentage relative error for the measurement of sensor capacitance.
The experimental results showing the pattern of error for different value of $R_x$ by varying $C_x$ from 10 pF to 42 pF is shown in Fig. \ref{fig:circuit_8}. The experimental results shown in Fig. \ref{fig:circuit_8} are obtained for Op-Amps LT1360 and LTC1049. 
Results show that the percentage relative error is high for the LT1360, which is also justified analytically. On the contrary, the error is small for LTC1049. The experimental result follows the analytical model of the circuit. The single-cycle averaging reduces the error to a great extent.\\ 
\begin{figure}
    \centering
    \includegraphics[trim=0 0 0 0, scale = 0.1]{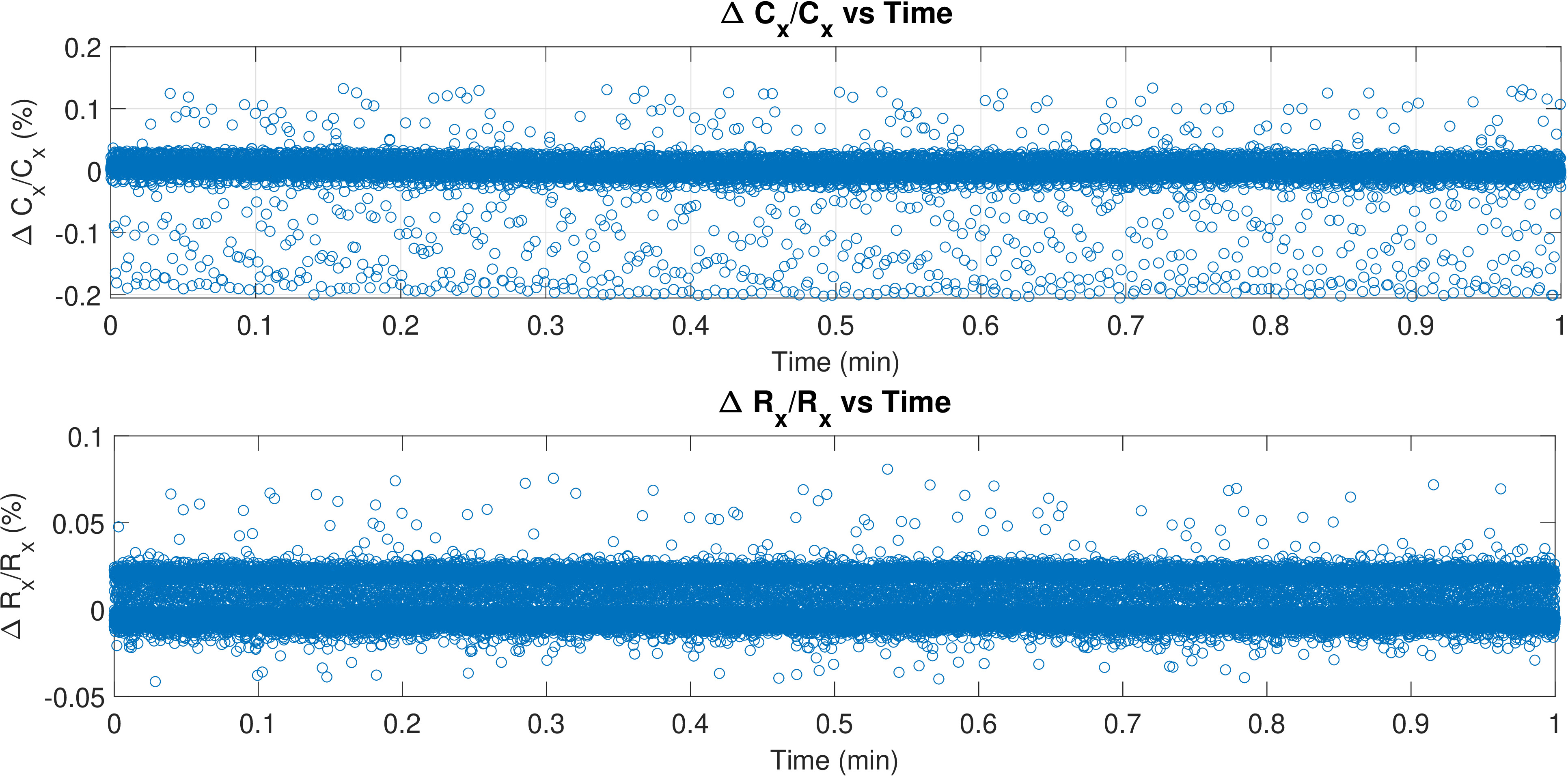}
     \caption{ Measured $\Delta C_x/C_x$ and $\Delta R_x/R_x$ with time, for $C_x$= 33 pF and $R_x$= 330 k$\Omega$}   
     \label{fig:circuit_counter}
\end{figure}
\textbf{Other Performance Parameters}\\
Other performance parameters such as the standard deviation and SNR of the proposed relaxation oscillator circuit for R-C sensors is measured experimentally. The time period $T_{p1}$, $T_{p2}$, $T_{p3}$, and $T_{p4}$ are experimentally recorded for a $R_x$ and $C_x$ values of 33 pF and 330 k$\Omega$. The measured periods are averaged for $C_x$ and $R_x$ measurement and the $\Delta C/C_x$  and $\Delta R/R_x$ is plotted in Fig. \ref{fig:circuit_counter}. The standard deviation and the SNR are calculated from the obtained data set using the expression mentioned in Table \ref{table:performace_parameters}. The obtained standard deviation and SNR for $C_x$ and $R_x$ are reported in Table \ref{table:performace_parameters}.

The circuit is tested for a capacitance as low as ten pF. At the lower capacitance value, the input capacitance of the operational amplifier is comparable with the sensor capacitance. Therefore, the relative error will be high at the lower capacitance value. However, the measurement can be performed by calibrating the effect of the input capacitance of the operational amplifier.
\begin{figure}[b]
    \centering
    \includegraphics[trim=10 33 15 15, scale = 1]{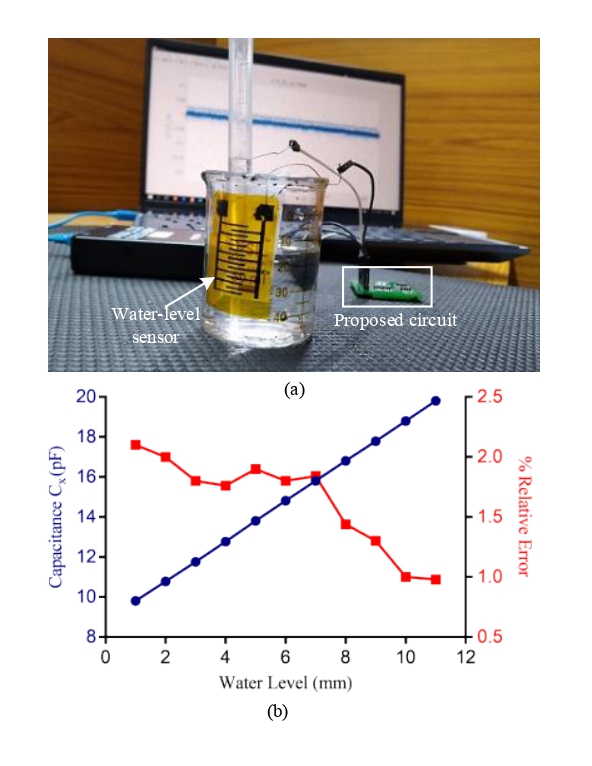}
     \caption{(a) Experimental setup for water-level leaky capacitive sensor, (b) experimental response of variation in the capacitance measured using the proposed circuit with multiple water levels. The sensor shows a 99.9\% linear response with respect to the water-level.}   
     \label{fig:water_level}
\end{figure}

\textbf{Testing with Water-Level Sensor}
The developed converter is utilized for the measurement of the sensor capacitor for the leaky capacitive water-level sensor. The water-level sensor is fabricated on a flexible substrate using the screen-printing technique. The sensor geometry is based on the inter-digitized electrodes. The principle of sensing is based on a change in the fringing field around the electrodes due to the water level, which in turn changes the capacitance of the sensor. The details of the sensor fabrication are reported in \cite{27,28}.

The experimental setup for the system is shown in Fig. \ref{fig:water_level}(a). The sensor is placed inside the beaker with known water levels. A controlled amount of water is poured into the beaker. The capacitance of the sensor is measured with the commercial LCR meter {(Agilent E4980A)} as well as with the proposed circuit. The percentage relative error in the measurement of the sensor capacitance for different water levels is shown in Fig. \ref{fig:water_level}(b).

The experimental test was conducted with the oscillator circuit built using the LTC1049 operational amplifier. In addition, the $R^2$ value is estimated from the fitted experimental data and found to be 99.9\%. This shows that the sensor shows a linear response to the water level. Further, the error reported in Table 2 is with the standard ceramic capacitors, while the error from the plot shown in Figure 11 is with the sensor. The mismatch in the error value is primarily because of the parasitic capacitance of the sensor-circuit interconnection and the measurement error.
\begin{table}
\caption{Performance Comparison of the Proposed Circuit with Other Reported Work}
\resizebox{\textwidth}{!}{
\begin{tabular}{|cc|cc|cc|cc|c|c|}
\hline
\multicolumn{2}{|c|}{\multirow{2}{*}{\textbf{Output}}}                                                                                                        & \multicolumn{2}{c|}{\textbf{Measurement Range}}   & \multicolumn{2}{c|}{\textbf{$\%$ Relative Error}} & \multicolumn{2}{c|}{\textbf{SNR(db)}} & \multirow{2}{*}{\textbf{Technique}} & \multirow{2}{*}{\textbf{\begin{tabular}[c]{@{}c@{}}Power\\ Consumption\end{tabular}}} \\ \cline{3-8}
\multicolumn{2}{|c|}{}                                                                                                                                        & \multicolumn{1}{c|}{$C_x$ pF} & $R_x$ k$\Omega$   & \multicolumn{1}{c|}{$C_x$}         & $R_x$        & \multicolumn{1}{c|}{$C_x$}   & $R_x$  &                                     &                                                                                       \\ \hline
\multicolumn{1}{|c|}{\textbf{\begin{tabular}[c]{@{}c@{}}This Work\\ (LTC1049)\end{tabular}}} & \textbf{\begin{tabular}[c]{@{}c@{}}Pulse\\ Width\end{tabular}} & \multicolumn{1}{c|}{10-42}    & 100-1000          & \multicolumn{1}{c|}{1.6}           & 0.62         & \multicolumn{1}{c|}{74.83}   & 78.96  & Oscillator                          & 8.4 mW                                                                                \\ \hline
\multicolumn{1}{|c|}{\textbf{\cite{29}}}                                                            & \textbf{Bit-Stream}                                            & \multicolumn{1}{c|}{100-200}  & 700-900           & \multicolumn{1}{c|}{$\pm$0.2}      & $\pm$0.2     & \multicolumn{1}{c|}{-}       & -      & Dual Slope                          & 175 mW                                                                                \\ \hline
\multicolumn{1}{|c|}{\textbf{\cite{30}}}                                                            & \textbf{Bit-Stream}                                            & \multicolumn{1}{c|}{50-1000}  & 50-1000           & \multicolumn{1}{c|}{0.73}          & 0.82         & \multicolumn{1}{c|}{64.35}   & 70.87  & Swithced-Capacitor                  & NA                                                                                    \\ \hline
\multicolumn{1}{|c|}{\textbf{\cite{15}}}                                                            & \textbf{Bit-Stream}                                            & \multicolumn{1}{c|}{150-206}  & \textgreater 5000 & \multicolumn{1}{c|}{1}             & -            & \multicolumn{1}{c|}{-}       & -      & Charge/Discharge                    & NA                                                                                    \\ \hline
\multicolumn{1}{|c|}{\textbf{\cite{20}}}                                                            & \textbf{Analog Voltage}                                        & \multicolumn{1}{c|}{100-2000} & 33-3000           & \multicolumn{1}{c|}{6}             & 8            & \multicolumn{1}{c|}{-}       & -      & PSD                                 & 168 mW                                                                                \\ \hline
\multicolumn{1}{|c|}{\textbf{\cite{18}}}                                                            & \textbf{Analog Voltage}                                        & \multicolumn{1}{c|}{10-760}   & 56-6500           & \multicolumn{1}{c|}{$\pm$0.11}     & $\pm$0.07    & \multicolumn{1}{c|}{49}      & 50     & Auto-Nulling                        & 142 mW                                                                                \\ \hline
\multicolumn{1}{|c|}{\textbf{\cite{19}}}                                                            & \textbf{Analog Voltage}                                        & \multicolumn{1}{c|}{200-100k} & 33-1200           & \multicolumn{1}{c|}{10}            & 100          & \multicolumn{1}{c|}{-}       & -      & Auto-Nulling                        & NA                                                                                    \\ \hline
\end{tabular}
}
\label{Table:Performance_comparison}
\end{table}

\textbf{Conclusion}\\
A modified relaxation oscillator based signal conditioning circuit for parallel R-C sensors is presented in this paper. The proposed circuit is based on a dual-slope and charge-transfer technique embedded into a modified relaxation oscillator. Detailed analysis and design criteria of the proposed circuit design are provided, considering various sources of errors. A single-cycle averaging method is analytically derived, which significantly reduces the effect of component non-idealities. The efficacy of the proposed circuit is tested on the fabricated prototype. The measurement result shows the effectiveness of the proposed circuit for parallel R-C sensors. The performance parameters of the developed prototype are calculated. The proposed circuit provides an SNR of around 75 dB and 79 dB for capacitive and resistive sensors, respectively. Overall, the proposed circuit is well suited for various sensors such as leaky capacitive sensors, resistive sensors with a parasitic capacitor, and impedance R-C sensors.
$\,$

$\,$

\end{document}